\documentclass[preprint,review,11pt]{elsarticle}
\usepackage{amssymb,amsmath}
\usepackage{graphicx}
\usepackage{epstopdf}
\usepackage{epsfig}
\usepackage{color}
\usepackage{lineno}
\usepackage{subcaption}
\usepackage{siunitx}
\usepackage{natbib}
\usepackage{lineno}
\usepackage{lscape}
\usepackage{multirow}

\usepackage[version=3]{mhchem}
\DeclareUnicodeCharacter{2061}{}
\DeclareUnicodeCharacter{0229}{}

\usepackage[figuresright]{rotating}

\usepackage[colorlinks=true,
            linkcolor=red,
            urlcolor=blue,
            citecolor=gray]{hyperref}

\journal{Journal of Hydrology}

\begin{document}
%

\begin{frontmatter}



\title{Multifractal detrended fluctuation analysis of rainfall time series in the Guadeloupe archipelago}

\author[cg]{Javier Gómez-Gómez}
\ead{f12gogoj@uco.es}

\author[ks,ua]{Thomas Plocoste\corref{cor1}}
\ead{thomas.plocoste@karusphere.com}

\author[lam,ha]{Esdra Alexis}
\ead{esdra.alexis@etu.univ-ag.fr}

\author[cg]{Francisco Jos{\'e} Jiménez-Hornero}
\ead{fjhornero@uco.es}

\author[cg]{Eduardo Guti{\'e}rrez de Rav{\'e}}
\ead{eduardo@uco.es}

\author[lam]{Silvere Paul Nuiro}
\ead{paul.nuiro@univ-antilles.fr}

\cortext[cor1]{Corresponding author}

\address[cg]{Complex Geometry, Patterns and Scaling in Natural and Human Phenomena (GEPENA) Research Group, University of Cordoba, Gregor Mendel Building (3rd ﬂoor), Campus Rabanales, 14071, Cordoba, Spain}

\address[ks]{Department of Research in Geoscience, KaruSphère Laboratory, Abymes 97139, Guadeloupe (F.W.I.), France}

\address[ua]{LaRGE Laboratoire de Recherche en Géosciences et Energies, Université des Antilles, F-97100 Pointe-à-Pitre, France}

\address[lam]{LAMIA Laboratoire de Mathématiques, Informatique et Applications, Université des Antilles, F-97100 Pointe-à-Pitre, France}

\address[ha]{Department of Mathematics (ENS), Université d’État d’Haïti (UEH), Port-au-Prince HT6110, Haïti}


\begin{abstract}

	Due to the vulnerability of the Caribbean islands to the climate change issue, it is important to investigate the behavior of rainfall. In addition, the soil of the French West Indies Islands has been contaminated by an organochlorine insecticide (Chlordecone) whose decontamination is mainly done by drainage water. Thus, it is crucial to investigate the fluctuations of rainfall in these complex environments. In this study, 19 daily rainfall series recorded in different stations of Guadeloupe archipelago from 2005 to 2014 were analyzed with the multifractal detrended fluctuation analysis (MF-DFA) method. The aim of this work is to characterize the long-range correlations and multifractal properties of the time series and to find geographical patterns over the three most important islands: Basse-Terre, Grande-Terre and Marie-Galante. This is the first study that addresses the analysis of multifractal properties of rainfall series in the Caribbean islands. This region is typically characterized by the almost constant influence of the trade winds and a high exposure to changes in the general atmospheric circulation. 12 stations exhibit two different power-law scaling regions in rainfall series, with distinct long-range correlations and multifractal properties for large and small scales. On the contrary, the rest of stations only show a single region of scales for relatively small scales. Hurst exponents reveal persistent long-range correlations which agree with other studies in nearby tropical locations. In the most eastern analyzed areas, larger scales exhibit higher persistence than smaller scales, which suggests a relationship between persistence and the highest exposure to the trade winds. Stronger conclusions can be drawn from multifractal spectra, which indicate that most rainfall series have a multifractal nature with higher complexity and degree of multifractality at the smallest scales. Furthermore, a clear dependence of multifractal nature on the latitude is revealed. All these results showed that MF-DFA is a robust tool to assess the nonlinear properties of environmental time series in a complex area.
	
\end{abstract}

\end{frontmatter}


\section{Introduction}	
\label{intro}

	Climate change is a worldwide issue \citep{fraga2012, melillo2014, ziervogel2014, chen2015}. In the literature, it is well known that Caribbean islands are among the most vulnerable countries on the planet \citep{field2012, stephenson2014, martinez2019, lowe2020}. This is often due to limited natural and human resources, smallness of the territories, urban and coastal areas densely populated and strong dependence on fragile sectors such as tourism and agriculture \citep{lewsey2004, scott2012, taylor2012, rhiney2015}. Over the past few decades, the frequency and intensity of extreme events in this area have steadily increased \citep{jennings2014, stephenson2014, stephenson2017, plocoste2022a}. Hydro-meteorological disasters such as droughts and floods are events that have significant socio-economic impacts on the development of these islands \citep{collymore2004, cashman2014, mycoo2018}. Between 1972 and 1993, the Chlordecone, an organochlorine insecticide, was spread  in banana fields of the French West Indies Islands \citep{coat2011, crabit2016}. The soil contamination by this pollutant will persist for several centuries because the only decontamination is through leaching by drainage water \citep{cabidoche2009}. To better quantify and predict the process of soil decontamination, it is crucial to investigate the fluctuations of rainfall in these complex environments.  

	Since the decade of 1950s, rainfall series and other natural phenomena are known to apparently exhibit long-range correlations \citep{hurst1951}. Since then, the study of correlations in the rainfall series and other signals of natural origin has emerged with some discrepancies in their determination which led to wide debates and the development of alternative methods for their computation \citep{mandelbrot1968, mandelbrot1969, wallis1970, klemevs1974, potter1976, peng1994, bisaglia1998}. More recently, different authors found the existence of fractal nature in rainfall series \citep{lovejoy1985, schertzer1987, sivakumar2001, maskey2016, dey2018, zhou2019, plocoste2021, jose2022}. However, later rainfall series were shown to be better characterized by different dimensions or fractal exponents, revealing their multifractal nature \citep{lovejoy1987, lovejoy1990, duncan1993}. In the last decades, different studies have been successful in characterizing these fractal exponents in relatively large records of rainfall series from distinct regions with the multifractal detrended fluctuation analysis (MF-DFA) \citep{kantelhardt2006, yu2014, baranowski2015, krzyszczak2017, tan2017, junior2018, adarsh2019, krzyszczak2019, zhang2019, adarsh2020, martinez2021, sarker2021, gomez2022, rahmani2022}. This method is suitable for multifractal analysis of time series because it computes the fluctuations for different statistical moment orders by previously removing trends in the series. Moreover, its results are comparable with those obtained by wavelet methods \citep{kantelhardt2002, kantelhardt2003}.
	
	In the Caribbean area, some studies have already modeled rainfall \citep{martis2002, ashby2005, gamble2008, stephenson2008, angeles2010}. Usually, they are based on statistical analyses. Knowing the nonlinear properties of rainfall as mentioned before, these models cannot take into account all of these fluctuations. Indeed, the results of these works are based only on a single time scale and might not necessarily reflect the features of rainfall series over several scales \citep{gomez2022}. It is therefore crucial to better understand the fluctuations of rainfall in these complex environments to subsequently develop prediction models. A widespread multifractal model is the generalized binomial multifractal model, which was already used by \cite{kantelhardt2006} to perform a fit of generalized Hurst exponents in rainfall and runoff records. The binomial multifractal series is defined as $x_k=a^{n(k-1)}b^{n_{max}-n(k-1)}$, with a total of $N=2^{n_{max}}$ numbers $k\,(k=1,...,N)$, where $n(k)$ is the number of digits equal to 1 in the binary representation of the index $k$ (see \cite{mali2015} for further details). In this model, the analytical expression for the fractal exponents or generalized Hurst exponents is $h(q)=1/q-[ln⁡(a^q+b^q)]/(q\,ln\,⁡2)$, while the width of the corresponding multifractal spectrum is $w=|ln\,⁡a-ln\,⁡b |/ln\,⁡2$. The main strength of this model is that it can be applied to both negative and positive moments $q$, unlike the Lovejoy and Schertzer’s multifractal model, which is only defined for positive moments (see \cite{kantelhardt2002, kantelhardt2003, kantelhardt2006} and references therein).	

	The main aim of this study is to characterize the spatial distribution of long-term correlations and relevant multifractal parameters of rainfall which can improve the understanding of the dynamic of water regime in the Caribbean basin and, more specifically, in the Guadeloupe archipelago. This is the first study to research geographical patterns of nonlinear and multifractal features in rainfall series in the Caribbean basin. The main objective of this work can be divided into the following steps: (i) to conduct an analysis of power spectra in 19 daily rainfall series; (ii) to perform a multifractal analysis with the MF-DFA method to obtain their multifractal properties and compare the results with power spectra; and (iii) to look for spatial patterns in the Guadeloupe archipelago by studying the relationships between the multifractal properties and the geographical features of gauge stations such as the orography, longitude, latitude and altitude. The section \ref{sitedata} of this study describes the experimental data and the study area; the section \ref{method} contains the methodology employed, including a summary of the MF-DFA method; the section \ref{results} includes the discussion of results; and the section \ref{conclusion} contains the main conclusions drawn from the analysis.

\section{Study area and data collection}	
\label{sitedata}

\begin{figure}[h!]
\begin{center}
\includegraphics[scale=0.75]{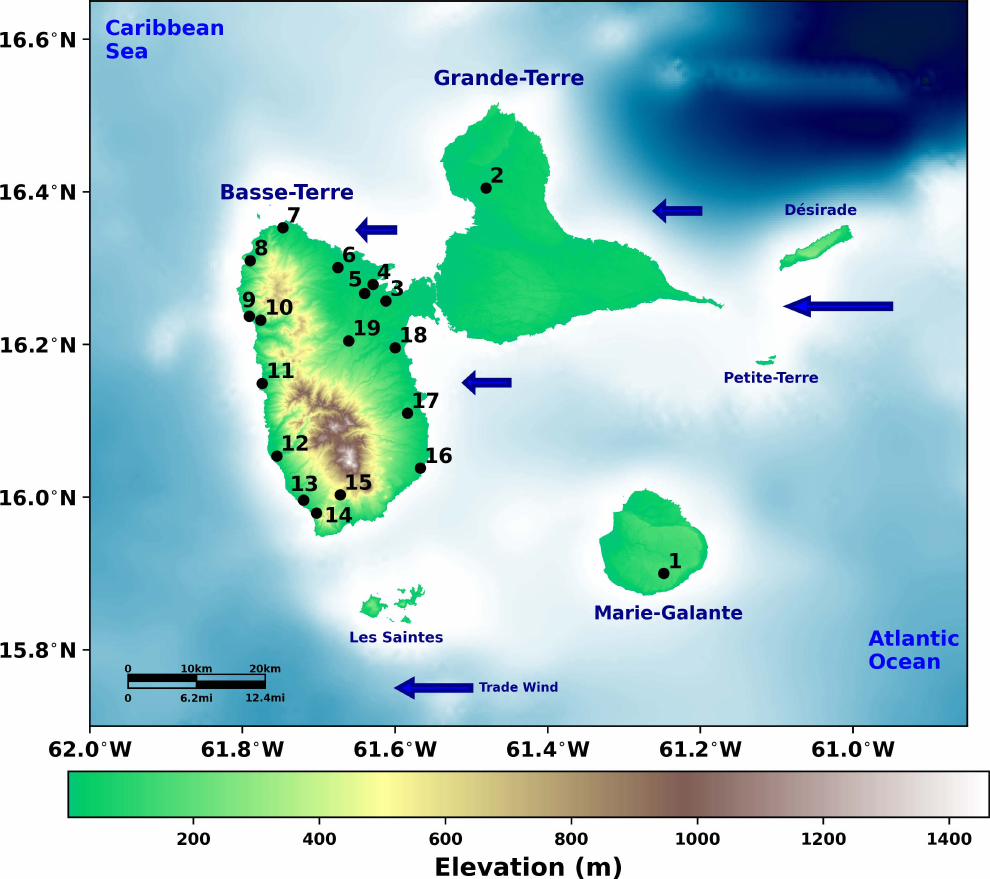}
\caption{\label{map} Map of the Guadeloupe archipelago with the locations of Météo France gauge stations available for this study. The arrows highlight the trade winds direction. }
\end{center}
\end{figure}

	To investigate the behavior of precipitation fluctuations in the Guadeloupe archipelago, ten years of daily rainfall is used. This Caribbean island is geographically located at latitude 16.25 north of the equator and longitude 61.58 west of the prime meridian. Situated in the middle of the Lesser Antilles, this French territory of 390,250 inhabitants has an area of $\sim$1800 $km^2$. Guadeloupe is made up of two main islands (Basse-Terre and Grande-Terre) and three small peripheral islands (La Désirade, Marie-Galante, and Les Saintes; see Figure \ref{map}) \citep{plocoste2018}. Grande-Terre is a sedimentary island with a low topography and a maximum elevation of 135 $m$ while Basse-Terre is a volcanic mountainous island with a maximum elevation of 1467 $m$. Like all the islands of the Caribbean area, Guadeloupe experiences a tropical rainforest climate (“Af”) according to the Köppen-Geiger classification \citep{peel2007, plocoste2022b}. The phenomena causing precipitation are divided into two categories \citep{brevignon2003}: advective phenomena and convective phenomena. Advection phenomena are the result of the movement of disturbances over the Atlantic such as tropical waves, cyclones, southern upwellings, i.e. large-scale events. On the other hand, the convective phenomena are the result of the creation of thermo-convective cells above a superheated surface: the hot air of the lower layers rises creating clouds, movement compensated by the collapse of cold air, i.e. mesoscale events. There are two main seasons: from January to June a dry and colder season and from July to December a rainy and hotter season \citep{plocoste2014}. Throughout the year, the study area is subject to the trade winds which are synoptic winds with an easterly component (90$^\circ$, see blue arrows in Figure \ref{map}) \citep{plocoste2020, plocoste2023}.
		
	To carry out this study, the rainfall time series of 19 gauges stations collected by Météo France network (https://meteofrance.gp/fr) using a Précis-Mecanique 3070 were used. The dataset available from January 2005 to December 2014 is complete with 3652 data points per time series. To ensure the quality of the data, all the time series have been pre-processed beforehand by Météo France. The spatial distribution of the stations is shown in Figure \ref{map} while their characteristics are presented in Table \ref{RGdistri}.

\begin{table}[h!]
\caption{\label{RGdistri} Description of gauge stations of daily rainfall database measured in the Guadeloupe archipelago from 2005 to 2014.}
\begin{center}
\small
\begin{tabular}{c c c c c c}
\hline
\multicolumn{1}{c}{City}& \multicolumn{1}{c}{Location} & \multicolumn{1}{c}{Station} & \multicolumn{1}{c}{Altitude ($m$)} & \multicolumn{1}{c}{Latitude ($^\circ$N)} & \multicolumn{1}{c}{Longitude ($^\circ$W)}\\
 \hline
Capesterre de Marie-Galante & Vidon & 1 & 146 & 15.900 & -61.248 \\
\hline
Petit-Canal & Godet & 2 & 35 & 16.405 & -61.481 \\
\hline
Baie-Mahault & Dupuy & 3 & 22 & 16.257 & -61.612 \\
\hline
\multirow{2}{*}{Lamentin} & Blachon & 4 & 16 & 16.279 & -61.629 \\
 & Bréfort & 5 & 35 & 16.267 & -61.640 \\
\hline
 \multirow{2}{*}{Sainte-Rose} & Galbas & 6 & 23 & 16.301 & -61.675 \\
 & Clugny & 7 & 10 & 16.353 & -61.747\\
\hline
Deshaies & Matouba & 8 & 42 & 16.310 & -61.790 \\
\hline
\multirow{2}{*}{Pointe-Noire} & Armand-Félix & 9 & 43 & 16.237 & -61.791 \\
 & Bellevue & 10 & 213 & 16.232 & -61.776\\
\hline
 Bouillante & Pigeon & 11 &34 & 16.149 & -61.774\\
\hline
 Vieux-Habitants & Beausoleil & 12 & 136 & 16.054 & -61.755\\
\hline
 Basse-Terre & Guillard & 13 & 92 & 15.996 & -61.720\\
\hline
\multirow{2}{*}{Gourbeyre} & Houelmont & 14 & 418 & 15.979 & -61.703\\
 & Dolé & 15 & 477 & 16.003 & -61.672\\
\hline
Capesterre Belle-Eau & Cayenne & 16 & 19 & 16.038 & -61.567 \\
\hline
Goyave & Christophe & 17 & 103 & 16.110 & -61.584\\
\hline
\multirow{2}{*}{Petit-Bourg} & Blonde & 18 & 55 & 16.196 & -61.600\\
 & Duclos & 19 & 110 & 16.205 & -61.661\\
  \hline
\end{tabular}
\end{center}
\end{table}

\section{Methodology}	
\label{method}

\subsection{Seasonality removal}	
\label{methodseason}

	Rainfall time series exhibit typical seasonal trends which can affect the nonlinear properties of these series \citep{kantelhardt2003, livina2011, gomez2022}. Thus, the seasonality must be filter out. In this work, seasonal trends were eliminated following the equation \citep{kantelhardt2006, junior2018, yang2019, gomez2022}:	

\begin{equation}
x^{'}_{i}=\frac{x_i-\mu _{i}}{\sigma_i}
\end{equation}
where $\mu _{i}$ is the calendar mean and $\sigma_i$ is the calendar standard deviation.

\begin{figure}[h!]
\begin{center}
\includegraphics[scale=0.9]{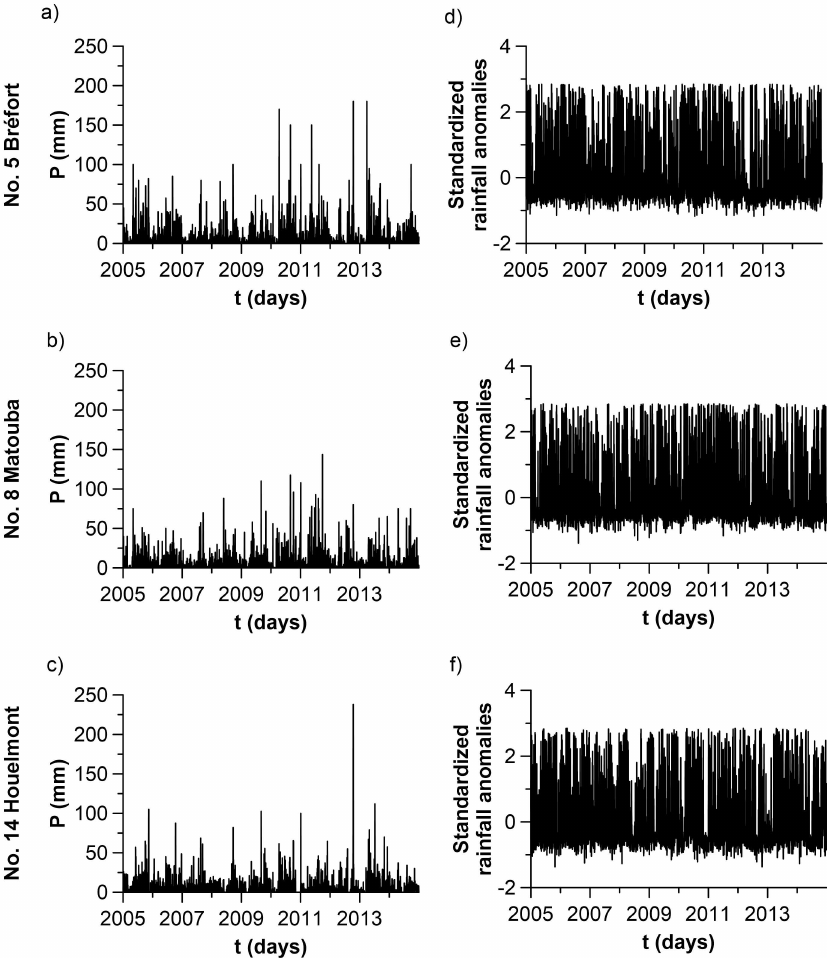}
\caption{\label{signal} (a-c) Original time series of rainfall of stations No. 5, 8 and 14, respectively. (d-f) Standardized rainfall anomalies of the same stations. }
\end{center}
\end{figure}

An example of these standardized rainfall anomalies computed for stations No. 5, 8 and 14 can be seen with their respective original series in Figure \ref{signal}.

\subsection{Power spectral density (PSD)}	
\label{methodPSD}

	Many physical processes in nature are described by stationary or non-stationary noisy time series which exhibit a PSD that follows a power-law asymptotic behavior for small frequencies, with $PSD(f)\sim f^{(-\beta)}$. A time series with $-1 < \beta <1$ can be modelled by a fractional Gaussian noise process (fGn), whereas one with $1 < \beta <3$ corresponds to a fractional Brownian motion (fBm) process. Furthermore, there exists a relation between these exponents and the standard Hurst exponent $H$ according to the following expressions \citep{eke2000, delignieres2006}:
	
\begin{equation}
H_{fGn}=\frac{\beta_{fGn} + 1}{2}
\label{PSDG}
\end{equation}

\begin{equation}
H_{fBm}=\frac{\beta_{fBm} - 1}{2}
\label{PSDB}
\end{equation}

The power spectra of the deseasonalized time series were performed to compare their results with those from the multifractal method improving the characterization of the signals.

\subsection{Multifractal detrended fluctuation analysis (MF-DFA)}	
\label{methodMFDFA}

	The MF-DFA method was applied to the deseasonalized rainfall time series. This method was developed by \cite{kantelhardt2002} as a generalization of the standard detrended fluctuation analysis (DFA) to characterize multifractal non-stationary time series. It consists of the following steps.
	
	Firstly, the “profile” is obtained subtracting the mean to every value in the series and computing the integral of the result. The “profile” is divided into $N_s$ non-overlapping segments of length $s$ from the beginning to the end of the series. The same procedure is repeated starting from the end to include the remaining part of the series. Thus, $2N_s$ segments are obtained for each scale $s$. The local trend $y_v (i)$ can be computed by a polynomial fit and subtracted from each segment $v$. Depending on the degree of the polynomial fit, the method is commonly named as MF-DFA1, MF-DFA2,... etc. In this work, a polynomial fit of order one was enough to eliminate trends in every case. Next, the variance $f^2(v,s)$ or the sum of squared differences between the values and trends in all segments is obtained \citep{kantelhardt2002}. Finally, the $qth$-order fluctuation function $F_q(s)$ is computed for different scales $s$ and statistical moment orders $q$:

\begin{equation}
F_q(s)= \left\{\frac{1}{2N_s} \sum^{2N_s}_{v=1} [f^2(v,s)]^{\frac{q}{2}} \right\}^{\frac{1}{q}}
\label{MFDFA1}
\end{equation}

The averaging procedure in Equation \ref{MFDFA1} must be replaced with a logarithmic averaging for $q=0$. See \cite{kantelhardt2002} for more details.

	The fluctuation function follows a power law $F_q(s)\sim s^{h(q)}$  if the analyzed time series is long-range power law correlated. Many climatic variables show long-range correlated fluctuations following a power law \citep{baranowski2015, baranowski2019, krzyszczak2019}. In particular, previous studies show that rainfall can exhibit different scaling regions which follow power laws with different exponents \citep{yang2019, gomez2022}.
	
	The exponent of the power law $h(q)$ is the generalized Hurst exponent. This exponent can be computed from the least-squares fit of log-log plots of $F_q(s)$ vs the scale for each $q$. Negative values of $q$ give more weight to small fluctuations, while positive ones give more weight to large fluctuations in the series. If $h(q)$ depends on $q$, the series has multifractal nature. On the contrary, if it is constant, the time series is monofractal \citep{kantelhardt2002, drozdz2015}. The standard Hurst exponent can be retrieved from the expression $H=h(2)-1$, if the analyzed signal is non-stationary, i.e., with a non-stationary variance, or from $H=h(2)$, if it is stationary, i.e., with stationary or constant variance \citep{eke2000}. This exponent has values which vary between 0 and 1 and allows to classify the behavior of long-range autocorrelations in the time series as persistent, if $H>0.5$, or antipersistent, if $H<0.5$ \citep{eke2002}.

	The fractal structure of the time series can also be described by the mass exponent $\tau(q)$, which is obtained from the expression $\tau(q)=qh(q)-1$ \citep{kantelhardt2002}. However, a more common descriptor of the multifractal strength is the multifractal spectrum $f(\alpha)$, which can be computed as follows:

\begin{equation}
f(\alpha)=q\alpha - \tau(q)
\end{equation}
where $\alpha=\tau^{'}(q)$ is the Hölder exponent or singularity strength.

	Different features were previously used in several studies to describe the multifractal spectra, such as the value of the Hölder exponent when $f(\alpha)$ is maximum, $\alpha_0$; the width, $w=\alpha_{max}-\alpha_{min}$ \citep{baranowski2019}; or the asymmetry \citep{drozdz2018, millan2022}. This study focuses on the analysis of $\alpha_0$ and $w$. The asymmetry is not considered here due to the omission of the negative statistical moments to avoid problems derived from the measurement precision, as it will be explained in section \ref{resultMFDFA}.

\section{Results and discussion}
\label{results}

\subsection{PSD analysis}	
\label{resultPSD}

\begin{figure}[h!]
\begin{center}
\includegraphics[scale=0.9]{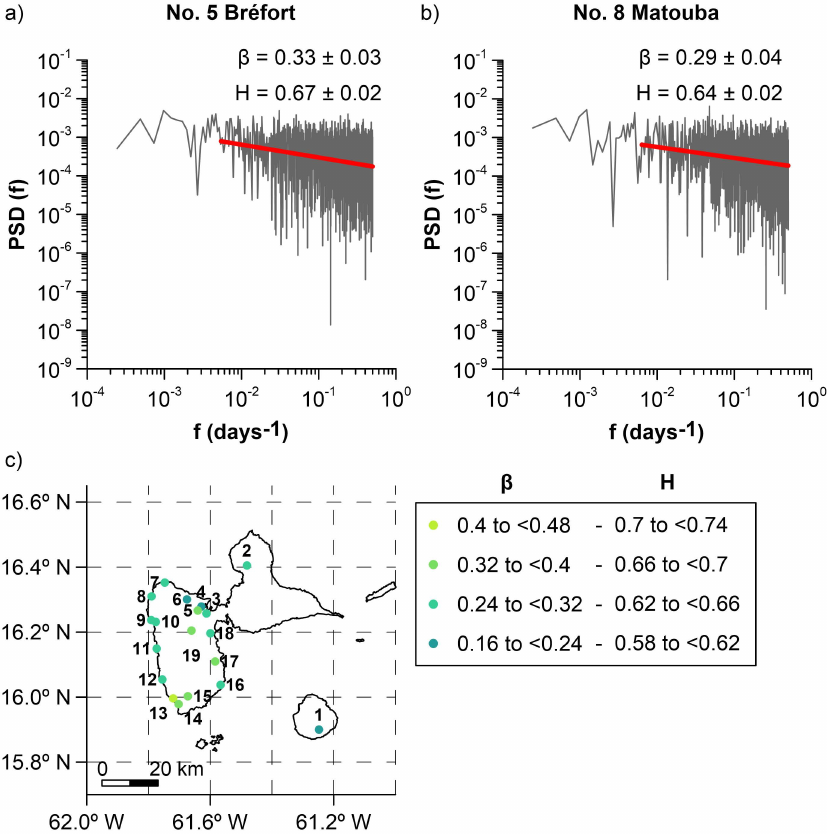}
\caption{\label{psd} (a-b) Log-log plots of the power spectral density (PSD) of stations No. 5 and 8 (grey lines) with their respective linear fits in the region of high frequencies (red lines). (c) Spatial distribution of $\beta$ and Hurst exponents computed from the PSDs. }
\end{center}
\end{figure}

	PSD was computed for every time series. Figures \ref{psd}a and b depict the PSD of stations No. 5 and 8. These stations exhibit a flat regime or spectral plateau for low frequencies (larges scales) and a power law behavior with PSD $(f) \sim f^{-\beta}$ for high frequencies (or small scales). Similar outcomes were obtained for the rest of stations, with different lower bounds for every case. Results of linear fits for large frequencies, including $\beta$ exponents with their errors, are shown in Table \ref{Tab2}. Low frequencies follow an uncorrelated process, whereas high frequencies follow a fractional Gaussian noise process with long-term correlations, which is typical of hydrological processes \citep{fraedrich1993, sivakumar2001, koscielny2006, sankaran2020, gomez2022}. Despite low Pearson correlation coefficients between PSDs and the frequency (below 0.3), all linear trends in log-log plots are statistically significant at the 1\% of significance level, as demonstrated by very low p-values (see Table \ref{Tab2}). Standard Hurst exponents were estimated from the PSD using the Equation \ref{PSDG}. Their values were found between 0.60 and 0.71, denoting that long-term correlations in these series are persistent.

\begin{table}[h!]
\caption{\label{Tab2} Results of linear fits of PSDs in the region of high frequencies. Power spectrum exponents ($\beta$), estimated standard Hurst exponents ($H$), Pearson correlation coefficients ($R$) and $p$-value of t-tests.}
\begin{center}
\small
\begin{tabular}{c c c c c}
\hline
\multicolumn{1}{c}{Station}& \multicolumn{1}{c}{$\beta$} & \multicolumn{1}{c}{$H$} & \multicolumn{1}{c}{$R$} & \multicolumn{1}{c}{$p$} \\
 \hline
1 & 0.21 $\pm$ 0.04 & 0.60 $\pm$ 0.02 & 0.10 & 4.03E-06\\
2 & 0.26 $\pm$ 0.04 & 0.63 $\pm$ 0.02 & 0.13 & 4.91E-09\\
3 & 0.27 $\pm$ 0.03 & 0.63 $\pm$ 0.02 & 0.17 & 7.15E-15\\
4 & 0.20 $\pm$ 0.04 & 0.60 $\pm$ 0.02 & 0.12 & 2.33E-08\\
5 & 0.33 $\pm$ 0.03 & 0.67 $\pm$ 0.02 & 0.21 & 9.39E-22\\
6 & 0.23 $\pm$ 0.04 & 0.61 $\pm$ 0.02 & 0.14 & 4.61E-10\\
7 & 0.27 $\pm$ 0.04 & 0.64 $\pm$ 0.02 & 0.17 & 2.19E-14\\
8 & 0.29 $\pm$ 0.04 & 0.64 $\pm$ 0.02 & 0.18 & 1.49E-15\\
9 & 0.31 $\pm$ 0.05 & 0.65 $\pm$ 0.02 & 0.15 & 6.76E-11\\
10 & 0.31 $\pm$ 0.04 & 0.65 $\pm$ 0.02 & 0.19 & 2.56E-17\\
11 & 0.27 $\pm$ 0.04 & 0.64 $\pm$ 0.02 & 0.16 & 2.07E-13\\
12 & 0.32 $\pm$ 0.04 & 0.66 $\pm$ 0.02 & 0.16 & 1.93E-12\\
13 & 0.41 $\pm$ 0.05 & 0.71 $\pm$ 0.02 & 0.20 & 4.39E-19\\
14 & 0.32 $\pm$ 0.03 & 0.66 $\pm$ 0.02 & 0.21 & 2.31E-21\\
15 & 0.37 $\pm$ 0.04 & 0.69 $\pm$ 0.02 & 0.23 & 1.32E-24\\
16 & 0.31 $\pm$ 0.03 & 0.66 $\pm$ 0.02 & 0.20 & 2.24E-19\\
17 & 0.33 $\pm$ 0.04 & 0.67 $\pm$ 0.02 & 0.17 & 2.19E-14\\
18 & 0.27 $\pm$ 0.04 & 0.64 $\pm$ 0.02 & 0.17 & 3.69E-14\\
19 & 0.33 $\pm$ 0.04 & 0.66 $\pm$ 0.02 & 0.17 & 6.88E-15\\
  \hline
\end{tabular}
\end{center}
\end{table}

	As all Caribbean islands, the Guadeloupe archipelago is characterized by specific climatic and topographic conditions, as commented in section \ref{sitedata}. Therefore, one must be cautious about comparing these results with others in different areas with a tropical climate. Globally, these results seem similar to those found in relatively close locations, such as in 8 of 10 stations in Venezuela, where annual rainfall was analyzed \citep{amaro2004}. They are also lower than those found for daily rainfall in the state of Tabasco, Mexico \citep{martinez2021}. Thus, the results are more similar to Venezuelan stations, which are in closer proximity to the Guadeloupe archipelago than the Tabasco state.

\begin{table}[h!]
\caption{\label{Tab3} Correlation coefficients between geographic characteristics of stations and their power and multifractal spectrum parameters. Hurst exponents are distinguished by method of estimation with a subscript. Superscripts stand for the regions (i) and (ii) in MF-DFA parameters. (*) refers to significant values at the 0.05 significance level and (**) refers to significant values at the 0.01 level.}
\begin{center}
\small
\begin{tabular}{c c c c}
\hline
\multicolumn{1}{c}{Parameter}& \multicolumn{1}{c}{Altitude} & \multicolumn{1}{c}{Latitude} & \multicolumn{1}{c}{Longitude} \\
 \hline
$\beta /H_{PSD}$ & 0.43 & -0.38 & -0.41\\
$H^{(i)}_{MF-DFA}$ & 0.20 & -0.45 & 0.08\\
$H^{(ii)}_{MF-DFA}$ & 0.39 & -0.08 & 0.14\\
$\Delta h^{(i)}$ & 0.24 & 0.46* & -0.21 \\
$\Delta h^{(ii)}$ & -0.13 & -0.40 & 0.12 \\
$\alpha^{(i)}_0$ & -0.06 & 0.03 & -0.13 \\
$\alpha^{(ii)}_0$ & 0.37 & -0.09 & 0.16\\
$w^{(i)}$ & -0.27 & 0.59** & -0.27 \\
$w^{(ii)}$ & 0.13 & -0.58* & 0.15\\
  \hline
\end{tabular}
\end{center}
\end{table}

	Table \ref{Tab3} shows the correlation coefficients between $\beta$ and $H$ exponents and the geographic features of stations, such as the altitude, latitude and longitude. Both parameters have the same values due to their linear relation shown in Equation \ref{PSDG}. Weak correlations with similar strength are exhibited by these exponents in the three cases. However, these results are not statistically significant at the 0.05 significance level of the correlation test. They contrast with the statistically significant correlations obtained for the latitude and altitude in the study of the Tabasco state, Mexico \citep{martinez2021}.

	Figure \ref{psd}c shows the spatial distribution of $\beta$ and $H$. It is shown a not very clear gradient for both parameters from the north-east to the south-west in the Base-Terre island, with similar intermediate values ($H$ between 0.62 and 0.66) in the western area. On the other hand, Marie-Galante and Grande-Terre islands (stations No. 1 and 2) display low and intermediate values, respectively. This means that the south-western area of the Base-Terre island exhibits some of the most correlated rainfall series. It must be highlighted that this region is the most hidden and protected from the trade winds and from general atmospheric circulation. This is due to its location at the south-west of the highest mountain peak in La Soufrière volcano (see Figure \ref{map}). Therefore, the orography seems to influence long-term correlations in rainfall series.

\subsection{MF-DFA analysis}	
\label{resultMFDFA}	

\begin{figure}[h!]
\begin{center}
\includegraphics[scale=0.6]{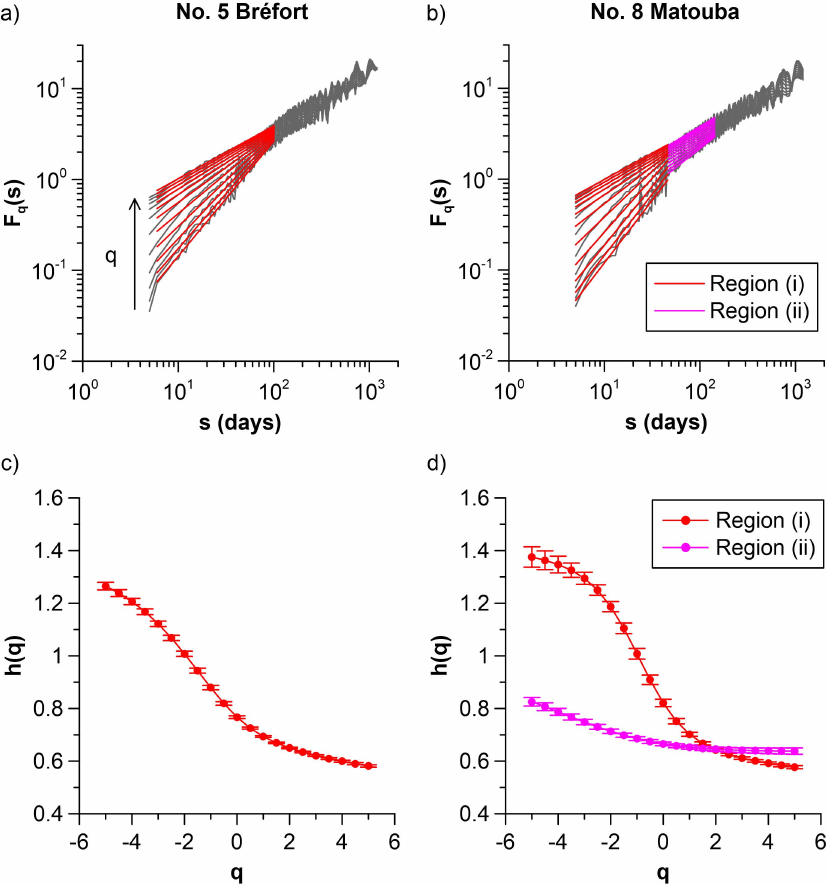}
\caption{\label{fig4} (a-s) Log-log plots of fluctuation functions vs scales. For clarity purposes, only statistical moments $q$=0, 2 and 5 are shown (green, blue and yellow dashed lines, respectively). Linear fits of log-log plots (red and magenta lines) are shown together with their corresponding lower and upper bounds. Two scaling regions, regions (i) (dark grey-shaded areas) and (ii) (light grey-shaded areas), can be observed in 12 stations, whereas the rest only exhibit a single scaling region with reliable linear fits.}
\end{center}
\end{figure}

	The MF-DFA method was applied to deseasonalized time series using scales $s$ from 5 to 1200 days with steps of 1 day and statistical moments $q$ in the range [0,5] with steps of 0.5. Negative statistical moments, which give more weight to small fluctuations, were discarded to avoid getting results biased by the magnification of measurement errors due to the limited precision of the measurement device. These errors can obscure the true small variations arising from the natural phenomenon. Figure \ref{fig4} shows the fluctuations functions for $q=0$, 2 and 5 for every station. As can be seen, the curves are ordered by increasing statistical moment $q$ from the bottom to the top. Two regimes with different exponents can be determined in 12 stations, where fluctuations fit well to a power law, one for small scales ($\sim$5-16 days to 1-2.5 months) and other for larger scales ($\sim$1-2.5 months to 5 months - 1 year). These regions were named as regions (i) and (ii), respectively. On the other hand, only 7 stations show one possible region, the region (i) (from 5-15 days to 2.5-6.5 months). Unreliable curves of $h(q)$ and multifractal spectra with non-concave shapes are obtained when attempting to fit the remainder of the curves of fluctuation functions at the largest scales. As a consequence, these linear fits were discarded in the analysis and evidences of only a single scaling region could be determined for these 7 stations. The lower and upper bounds of linear fits were different in every case. Moreover, these bounds do not match the ones obtained in linear fits of log-log plots of PSDs. Pearson correlation coefficients between the fluctuation functions and the scale in the log-log plots were between 0.97 and 1.00. The authors assume that micro-climates may be responsible for regime differences between stations.

\begin{figure}[h!]
\begin{center}
\includegraphics[scale=0.7]{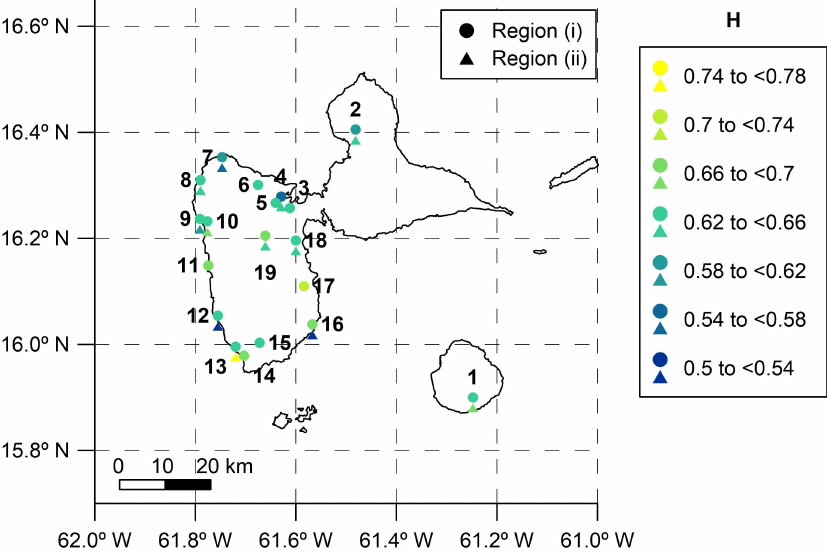}
\caption{\label{fig5} (a-l) Generalized Hurst exponents $h(q)$ and their statistical errors of stations with two different scaling regions: regions (i) (red circles) and (ii) (magenta squares). Dashed lines stand for a constant value of $h=0.5$ or a monofractal uncorrelated process. }
\end{center}
\end{figure}

\begin{figure}[h!]
\begin{center}
\includegraphics[scale=0.7]{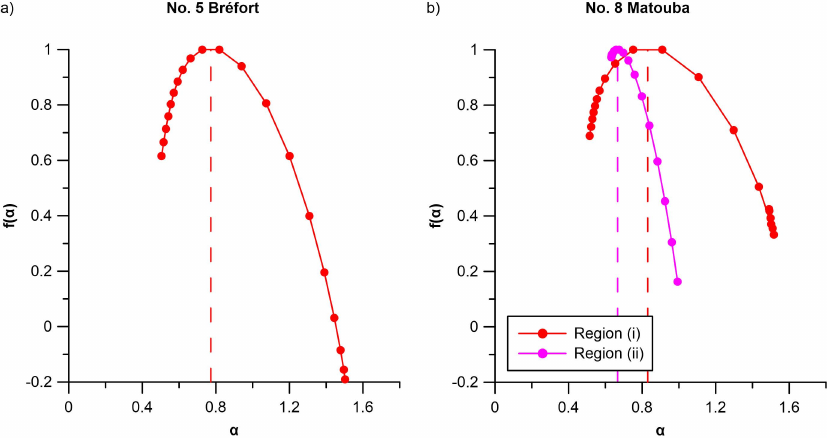}
\caption{\label{fig6} (a-g) Generalized Hurst exponents $h(q)$ and their statistical errors of stations with a single scaling region (red circles). }
\end{center}
\end{figure}

	Generalized Hurst exponents $(h(q))$ of stations which show two significant scaling regions are depicted in Figure \ref{fig5}. Statistical errors are 0.02 or lower in every case. These exponents decrease with $q$ in both cases for regions (i) and (ii), denoting that these time series might be multifractal. $h(q)$ exhibits a lower range of exponents for larger scales (region (ii)), with $\Delta h$ between 0.03 and 0.09, whereas the region (i) exhibits a range of exponents between 0.08 and 0.35. This suggests that rainfall series could exhibit multifractality at small scales whereas they might be monofractal or have a weaker multifractality at larger scales. These results seem to highlight the two origins of rainfall on the island, i.e., convective phenomena (mesoscale events) and advective phenomena (large scale events). The slightly higher uncertainties on $h(q)$ in region (i) might be due to that rainfall related to the mesoscale events is more intermittent. The time scales of the region (ii) emphasize the seasonal behavior of rainfall in the Caribbean area, e.g., the hurricane season from June to October \citep{tartaglione2003, dunion2011}. On the other hand, the other 7 stations with a single scaling region (region (i)) are depicted in Figure \ref{fig6}. Statistical errors are 0.01 or lower in every case. In these stations, $\Delta h$ is between 0.07 and 0.19.

\begin{figure}[h!]
\begin{center}
\includegraphics[scale=0.7]{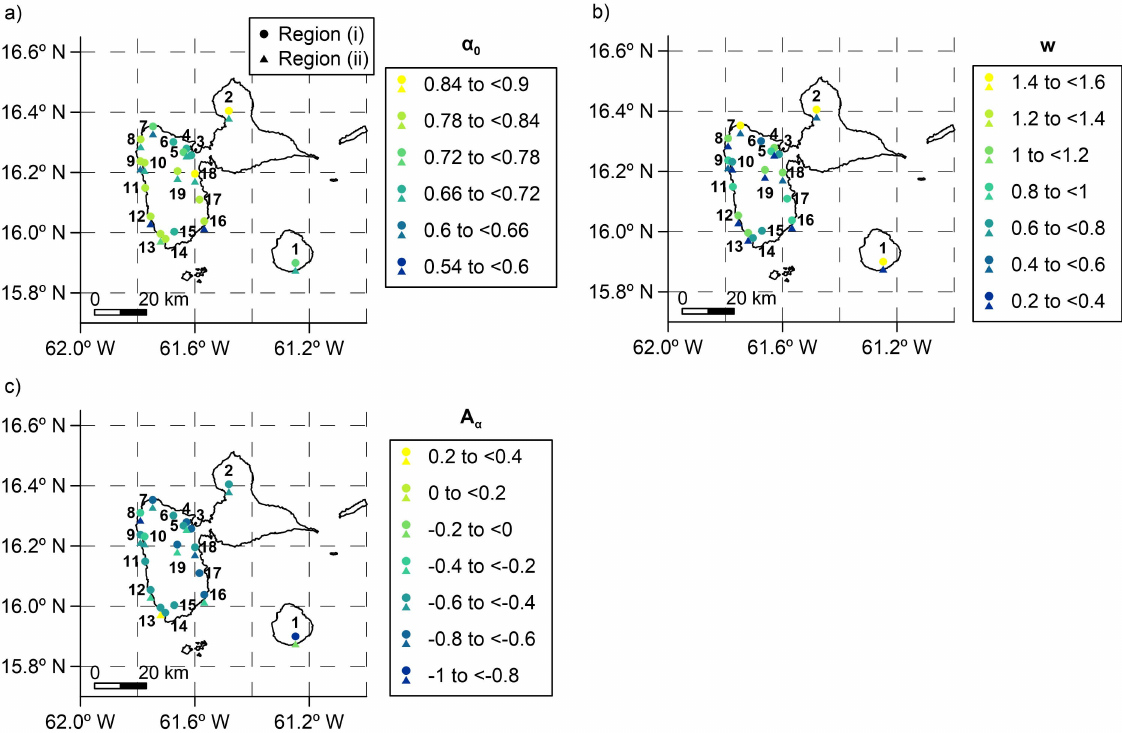}
\caption{\label{fig7} (a) Spatial distribution of the Hurst exponents obtained from the MF-DFA method for regions (i) (circles) and (ii) (triangles). (b) Spatial distribution of the range of fractal exponents ($\Delta h$) for both scaling regions. }
\end{center}
\end{figure}

	The spatial distribution of the standard Hurst exponent $H$, i.e., $h(2)$, is depicted in Figure \ref{fig7}a for all stations with only one or with two scaling regimes. Stations with only one scaling region are marked with a colored circle (region (i)), while those with two scaling regions are marked with a circle (region (i)) and a triangle (region (ii)). Both scaling regions do not display statistically significant correlations between the Hurst exponents and the geographic features as in the case of the PSD results (see Table \ref{Tab3}), and contrary to the outcomes found in the Tabasco state, Mexico \citep{martinez2021}. Moreover, all values display a wider range than in the case of the PSD (between 0.5 and 0.75), although the similarities with Venezuelan stations are retained \citep{amaro2004}. For region (i), the distribution of values (with $H$ between 0.57 and 0.70) is heterogeneous and denotes persistence in these time series. Although stations in Marie-Galante and Grande-Terre islands (stations No. 1 and 2) show different values, these results have some similarities with exponents shown in Figure \ref{psd}c, such as the lowest persistence located at the north of the Basse-Terre island and the abundance of values between 0.62 and 0.66. The discrepancy in results between PSD and MF-DFA methods seems to be due to strong fluctuations of PSDs, producing unreliable linear fits. The region (ii) of larger scales found in 12 stations exhibits even more heterogeneity in their spatial distribution and a wider range of $H$ (between 0.5 and 0.75). Stations No. 12 and 16 exhibit the lowest values in the scaling region (ii) (0.54 and 0.51, respectively), denoting an almost uncorrelated behavior in larger scales, which suggests a relation to the spectral plateau found in the region of lower frequencies in the PSDs for these stations. On the other hand, the station No. 13 shows the highest value, in the same way as in the results of PSDs (see Table \ref{Tab2}). Stations No. 1 and 2 in Marie-Galante and Grande-Terre islands show higher persistence in region (ii) than in region (i), probably due to the high exposure to the trade winds and the general atmospheric circulation which mainly influence the larger scales.
	
	The spatial distribution of $\Delta h$ is shown in Figure \ref{fig7}b. These results exhibit a heterogeneous distribution. However, the highest values in the region (i) are more likely to be found to the north and the northwest, whereas lower values are more likely to be found to the south and the southeast. On the contrary, as already mentioned above, the range of exponents for region (ii) is very narrow. Correlation tests show statistically significant correlations between this variable and the latitude only in region (i) of scales, where there is a larger range of values (see Table \ref{Tab3}). These outcomes indicate that the higher the latitude, the higher the degree of the multifractality is at small scales.
	
\begin{figure}[h!]
\begin{center}
\includegraphics[scale=0.5]{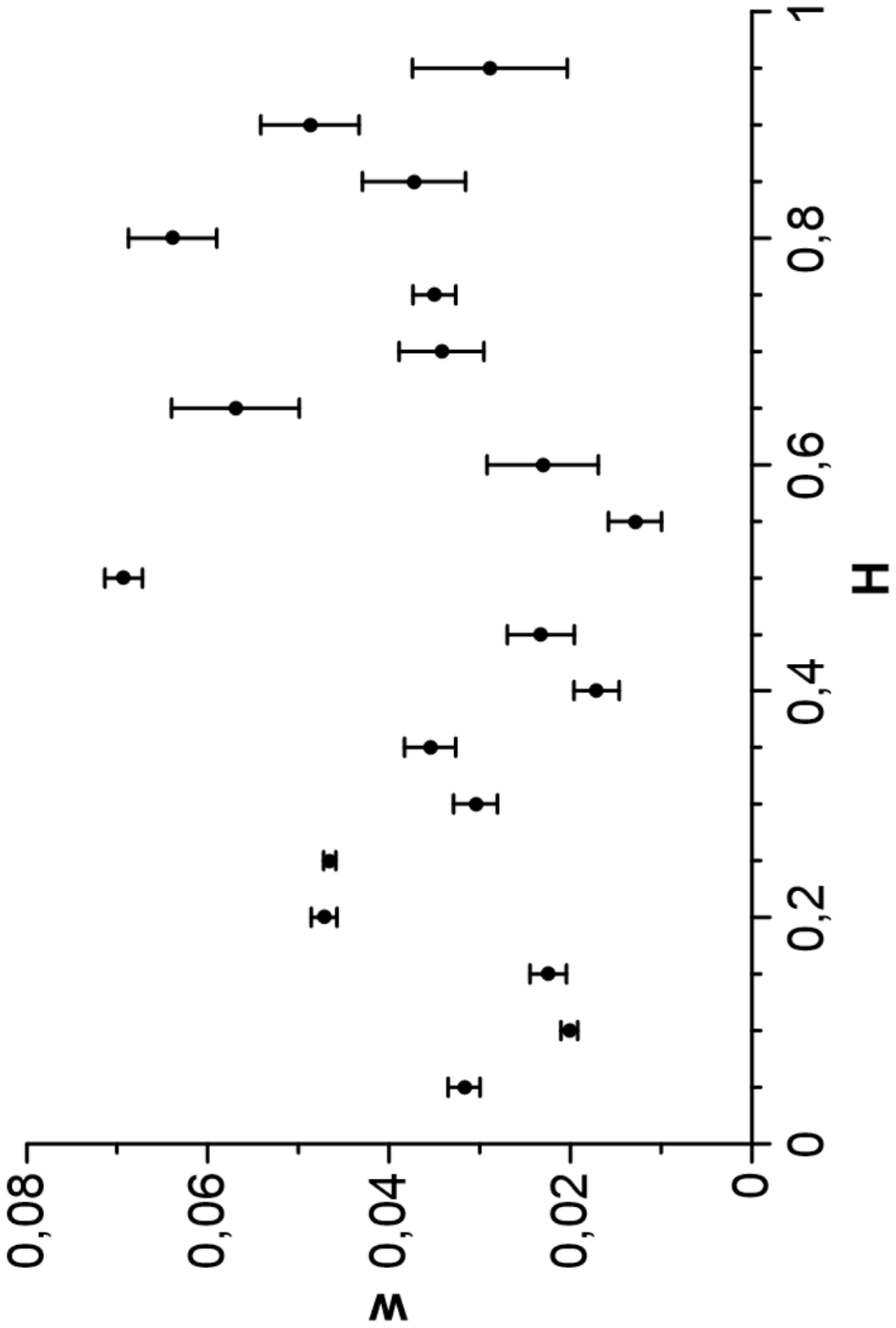}
\caption{\label{fig8} Average width of multifractal spectra ($w$) and 95\% confidence intervals of simulated fGn series arranged by their theoretical Hurst exponent ($H$). 30 samples of the same length as the real time series were generated for each value of the Hurst exponent in the range (0,1) with steps of 0.05. The MF-DFA method was used with the same scales (from 5 to 1200 days with steps of 1 day) and $q$ values ([0,5] with steps of 0.5) as in rainfall series.}
\end{center}
\end{figure}

	Concerning the multifractal spectrum analysis, when the MF-DFA method is used on monofractal time series their multifractal spectra can display a low non-zero value of width. Therefore, to properly differentiate these signals by their fractal nature, an experiment was performed. 330 simulated signals of fGn processes with different Hurst exponents were analyzed with the MF-DFA method using the same values for the scales and statistical moments as in the real time series. Hurst exponent values were chosen in the range [0.05,0.95] with steps of 0.05. For each value of $H$, 30 samples were artificially generated with the same length as the rainfall series. The width of the multifractal spectrum was obtained for every signal. Next, the average and 95\% confidence intervals were computed for each value of $H$. The results can be seen in Figure \ref{fig8}. For every value of $H$, confidence intervals are below 0.075. As a consequence, this value was chosen as a threshold value to differentiate between monofractal and multifractal series. When a multifractal spectrum exhibits a width of 0.075 or higher, the corresponding time series is considered as multifractal. This criterion is valid for time series with similar features (long-term correlations, length…) as those studied here, for which only the statistical moments $q \geq 0$ can be considered.	

\begin{figure}[h!]
\begin{center}
\includegraphics[scale=0.7]{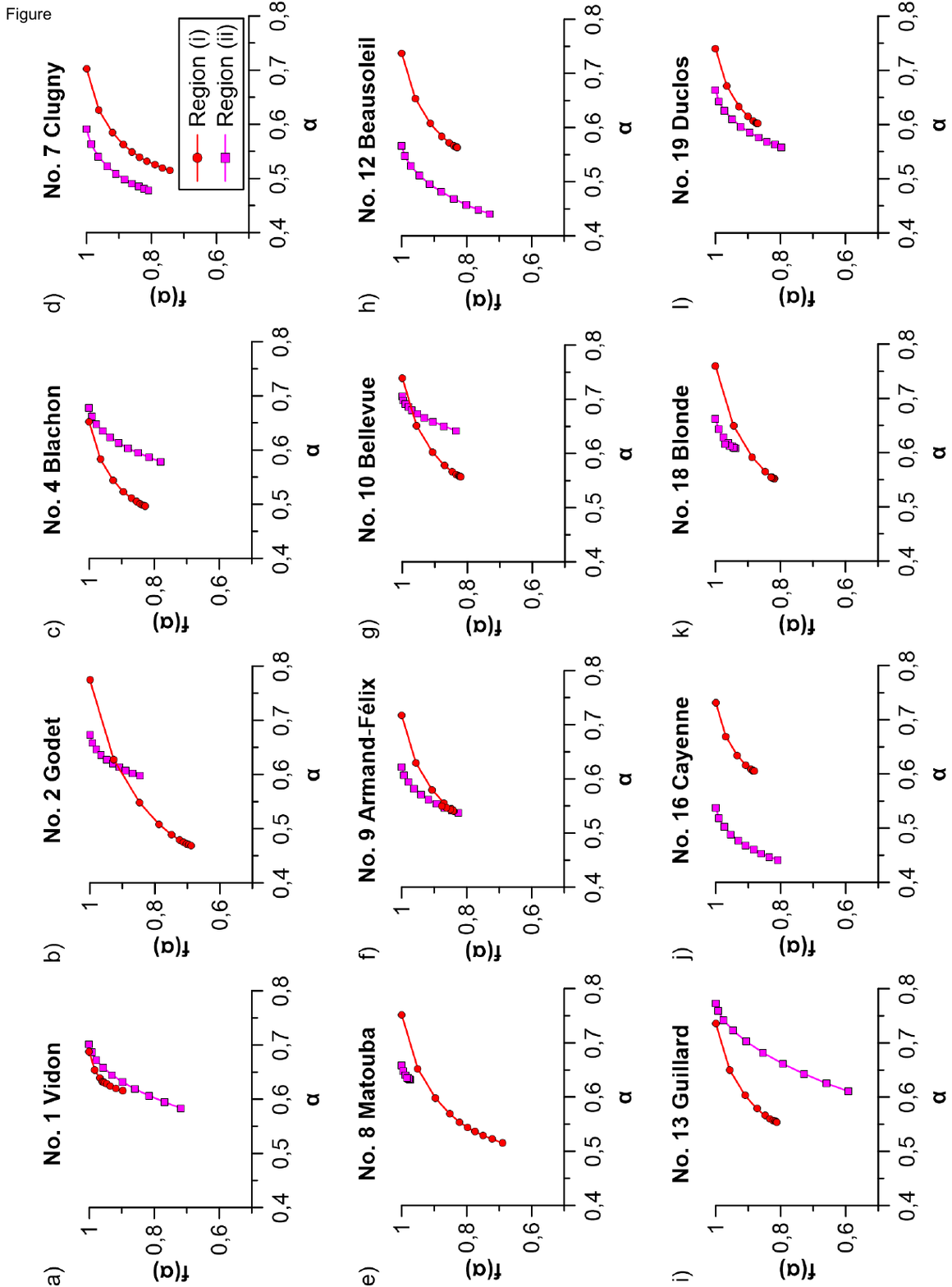}
\caption{\label{fig9} (a-l) Multifractal spectra of stations with two different scaling regions: regions (i) (red circles) and (ii) (magenta squares).}
\end{center}
\end{figure}

	Figure \ref{fig9} depicts multifractal spectra of stations with two significant scaling regions. As can be seen in the figure, multifractal spectra from region (i) are more similar in most stations, except in the case of stations No. 2 and 8 which have a broader width, denoting a stronger multifractal behavior \citep{baranowski2019, gomez2022}, and station No. 1, with a significantly narrower width. On the contrary, the region (ii) of larger scales exhibits very different spectra, with the time series recorded in stations No. 8 and 18 being almost monofractal. This phenomenon will be addressed in more detail in the analysis of the spatial distribution.	

\begin{figure}[h!]
\begin{center}
\includegraphics[scale=0.7]{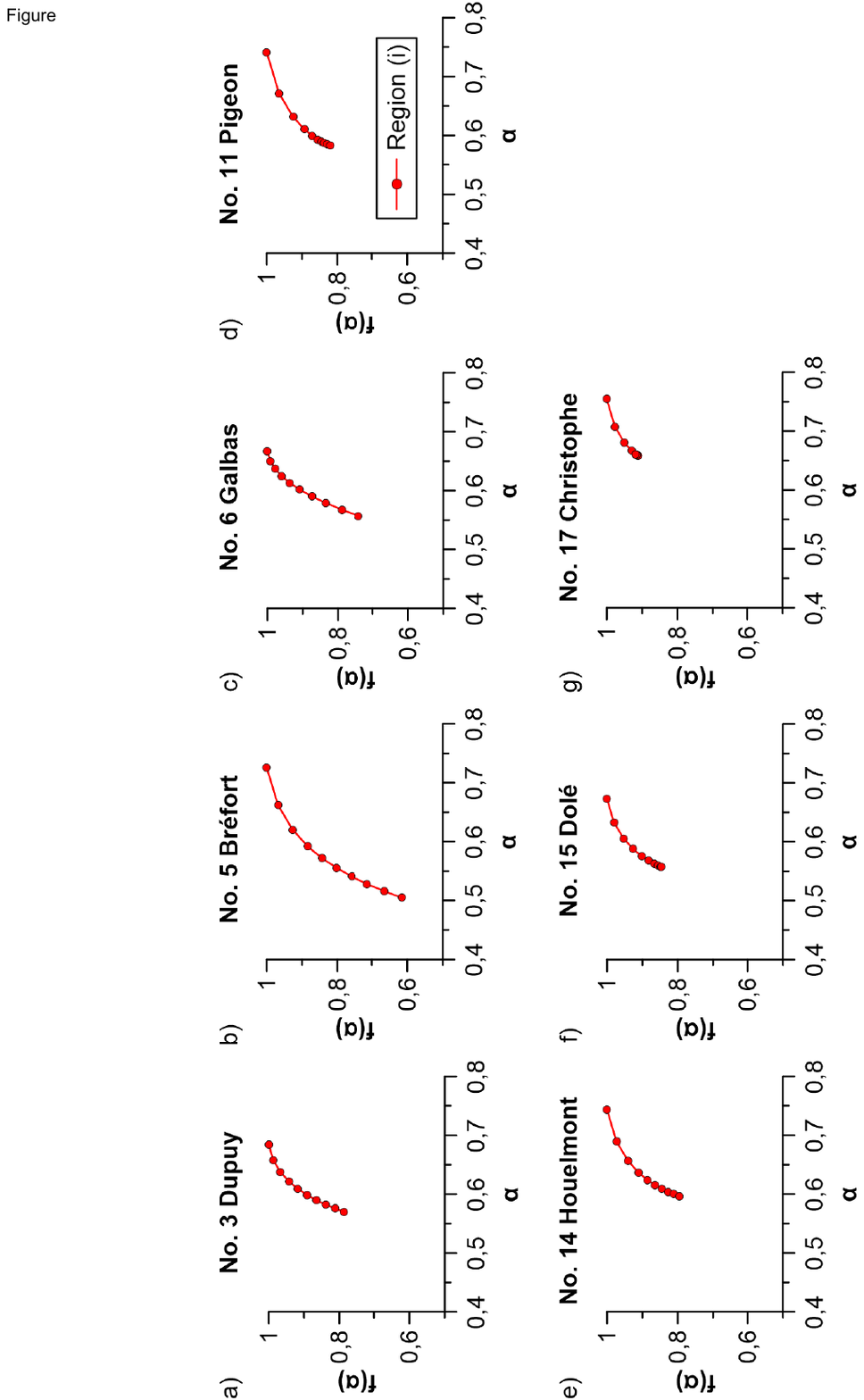}
\caption{\label{fig10} (a-g) Multifractal spectra of stations with a single scaling region (red circles).}
\end{center}
\end{figure}

	Multifractal spectra of stations with a single scaling region are shown in Figure \ref{fig10}. They show similar values of $\alpha_0$ and different values of width. The station No. 5 stands out for a wider spectrum and a more complex time series.

\begin{figure}[h!]
\begin{center}
\includegraphics[scale=0.7]{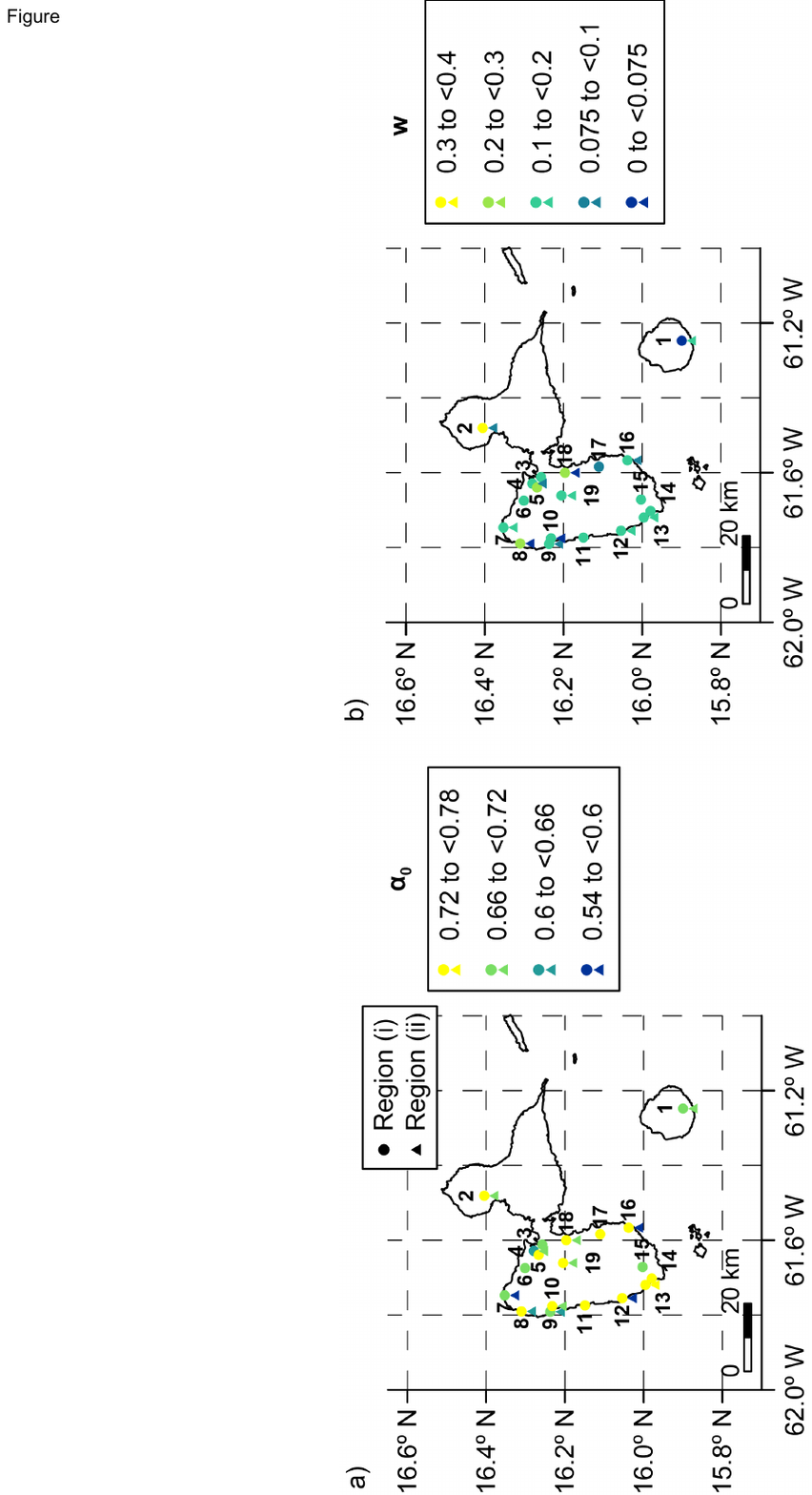}
\caption{\label{fig11} Spatial distribution of multifractal spectrum parameters for regions (i) (circles) and (ii) (triangles). (a) The singularity exponent at the maximum spectrum ($\alpha_0$). (b) The width ($w$).}
\end{center}
\end{figure}

	The spatial distribution of the analyzed parameters of multifractal spectra ($\alpha_0$ and $w$) are shown in Figure \ref{fig11}. Figure \ref{fig11}a depicts the distribution of $\alpha_0$, where region (i) displays a homogeneous area which cover the most part of the mid and southern area of the Basse-Terre island with values between 0.72 and 0.78. The Grande-Terre station (No. 2) also shows a similar value. Lower values ($\alpha_0$ between 0.6 and 0.72) are exhibited in the northern and northwestern areas, in station No. 15, and in the Marie-Galante station (No. 1). The region (ii) shows similar or lower values than the region (i) in the 12 stations with two regimes, denoting that some stations exhibit lower complexity at larger scales. Therefore rainfall time series in these locations have fine structure and the most complexity at the smallest scales \citep{baranowski2019}. Stations No. 12 and 16 show the lowest values in the same way as results of Hurst exponents, which seems to be due to the lower correlations and a more random behavior in their respective series. Correlations found between $\alpha_0$ and the geographic features of the stations were not statistically significant in any scaling region (see Table \ref{Tab3}).
	
	The results of $w$ are shown in Figure \ref{fig11}b. The width is an indicator of the degree of multifractality of a signal \citep{krzyszczak2019}. Most stations in the Basse-Terre island exhibit a width in region (i) between 0.1 and 0.2, while three stations show higher values and only one station has a narrower width. The Grande-Terre station (No. 2) displays the widest spectrum (0.306) and the Marie-Galante station (No. 1) shows the narrowest one (0.072). According to what it was discussed above, the station No. 1 exhibits a monofractal nature at small scales. The rest of the stations display a multifractal behavior in rainfall series, with the station No. 2 showing the highest degree of multifractality and the widest range of singularity exponents. Larger scales (region (ii)) show more heterogeneity. The narrowest spectra are found in stations No. 8, 10 and 18 (with w between 0 and 0.075), denoting that these rainfall series are multifractal at small scales but monofractal at larger scales. On the contrary, the Marie-Galante station (No. 1) exhibits the opposite behavior, with a monofractal nature at small scales but multifractal at larger scales. Despite their differences in the spatial distribution, both regions of scales display significant correlations with the latitude but of opposite sign, as it can be seen in Table \ref{Tab3}. Thus, a higher latitude significantly influence the presence of a higher degree of multifractality in rainfall time series in Guadeloupe at small scales (from several days to some months) and less multifractality at large scales (from several months to one year). This phenomenon translates into a high dependence of the rainfall complexity on the geographic location.

\section{Conclusion}
\label{conclusion}

	The aim of this paper was to investigate the fluctuations of rainfall in an insular context using a multifractal framework. To carry out this study, 19 rainfall series in the Guadeloupe archipelago were used. All 19 rainfall series reveal two different regimes of frequencies in their power spectra. One regimen displays a flat uncorrelated spectrum for low frequencies and the other exhibits power-law long range persistent correlations for high frequencies. The main conclusion drawn in the analysis of fluctuations functions is that 12 stations exhibit rainfall series with two different scaling regions, with distinct different multifractal properties on large and small scales. The smallest scales cover from several days to some few months, while the largest scales include from several months to one year. On the contrary, the rest of 7 stations display a single region with a power-law behavior from several days to several months. The discrepancy in the range of scales and the strength of long-range correlations with the PSD analysis might be derived from unstable fluctuating power spectra that yielded unreliable results.
	
	Hurst exponents are similar to those found in other nearby tropical areas. Furthermore, results of MF-DFA suggest that larger scales exhibit higher persistence than smaller scales in the most eastern analyzed areas, in Grande-Terre and Marie-Galante islands. This fact suggests a relation between the higher exposure to the trade winds and higher long-range correlations in rainfall series.
	
	More relevant conclusions can be drawn from the features of multifractal spectra. It can be confirmed that rainfall series are more complex and have fine structure in most series at the smallest scales, whereas lower complexity is exhibited at larger scales. Furthermore, most stations display a multifractal nature with a spectral width larger than a threshold value of 0.075. Only 4 stations, the station No. 1 at small scales and the stations No. 8, 10, 18 at large scales, show a monofractal behavior. The most degree of multifractality and the most complexity are found at the smallest scales in the Grande-Terre station (No. 2). Larger scales exhibit the same or lower degree of multifractality, except in the Marie-Galante station (No. 1). Statistically significant correlations with the latitude of opposite signs were found for both scaling regions. Thus, there is a clear dependence of multifractal nature on the geographic location in rainfall series in the Guadeloupe archipelago. All these results clearly demonstrated the impact of rainfall origins, i.e., from mesoscale to large scale events, on their degree of fluctuation. Despite the promiscuity of certain stations, different regimes are highlighted showing the impact of micro-climates.

	The multifractal characterization of rainfall series in the Guadeloupe archipelago improves the understanding of the underlying nonlinear relationships between the temporal organization of the intensity of daily rainfall and the general atmospheric circulation and the geographical location in this area, with a clear dependence on the latitude. Moreover, the conclusions drawn in this study can be relevant for future studies that address the description of nonlinear features of climate or the improvement of climatic predictive models in the Caribbean basin.

\section*{Acknowledgements}

The authors gratefully acknowledge the support of the funding sources. The authors would like to thank the French Met Office (M\'et\'eo France Guadeloupe) for providing meteorological data. The authors also acknowledge Mr. Sylvio Laventure (geomatics specialist) for his cartographic assistance and the anonymous reviewers for their valuable comments and constructive suggestions.

\section*{Disclosure statement}

No potential conflict of interest was reported by the authors.

\section*{Funding}

This research was funded by the European Regional Development Fund (Research project UCO-1379178, Operational Program Framework Andalusia 2014–2020) and the Andalusian Research Plan Group TEP-957. 

\clearpage

\bibliographystyle{model4-names}
\biboptions{authoryear}

\bibliography{RainMFDFA}

\end{document}